\documentclass[review,12pt,authoryear]{elsarticle}
\usepackage[utf8]{inputenc} % use source file containing diacritics

\usepackage{graphicx}
\usepackage{textcomp}
\usepackage{subfigure} %AGREGADO PARA EL SUBFIGURE
\usepackage{amssymb}
\usepackage{amsmath}
\usepackage{setspace}

\usepackage{cite}
\journal{Advances in Space Research}

\begin{document}

\begin{frontmatter}


\title{Near-real-time VTEC maps: new contribution for Latin America Space Weather}




\author[unlp,conicet]{Luciano P. O. Mendoza}
\author[unlp,conicet]{Amalia M. Meza}
\author[unlp,conicet]{Juan Manuel Arag\'on Paz}

\address[unlp]{Laboratorio de Meteorolog\'\i{}a espacial, Atm\'osfera terrestre, Geodesia, Geodin\'amica, dise\~no de Instrumental y Astrometr\'\i{}a (MAGGIA), Facultad de Ciencias Astron\'omicas y Geof\'\i{}sicas (FCAG),
Universidad Nacional de La Plata (UNLP), Paseo del
Bosque s/n, B1900FWA, La Plata, Argentina}

\address[conicet]{Consejo Nacional de Investigaciones Cient\'\i{}ficas y T\'ecnicas (CONICET), Godoy Cruz 2290, C1425FQB, Buenos Aires, Argentina}

\begin{abstract}
The development of regional services able to provide ionospheric vertical total electron content (VTEC) maps and ionospheric indexes with a high spatial resolution, and in near-real-time, are of great importance for both civilian applications and the research community. We provide here the methodologies, and an assessment, of such a system. It relies on the public Global Navigational Satellite Systems (GNSS) infrastructure in South America, incorporates data from multiple constellations (currently GPS, GLONASS, Galileo and BeiDou), employs multiple frequencies, and produces continental wide VTEC maps with a latency of just a few minutes. To assess the ability of our system to model the ionospheric behavior we performed a year-round intercomparison between our near‐real‐time regional VTEC maps, and VTEC maps of verified quality produced by several referent analysis centers, resulting in mean biases lower than 1 TEC units (TECU). Also, the evaluation of our products against direct and independent GNSS-based slant TEC measurements shows RMS values better than 1 TECU. In turn, ionospheric weather W-index maps were generated, for calm and disturbed geomagnetic scenarios, solely employing our quality verified VTEC maps. The spatial representation of these W-index maps reflects the state of the ionosphere, with a resolution of $ 0.5\times 0.5$ degrees. Finally, we conclude that our products, computed every 15 minutes, do provide an excellent spatial representation of the regional TEC, and are able to provide the bases for the possible computation of ionospheric W-index maps, also in near-real-time.
\end{abstract}


\end{frontmatter}

%%\linenumbers

\section{Introduction}
\label{S:1}

The requirement for near-real-time products based upon current ionospheric behavior has led to focus the interest of the scientific community on the real-time ionospheric monitoring. These products are required both in scientific applications  and in practical services. Among others, for the monitoring of space weather events such as solar flares, solar energetic particles events and coronal mass ejections \citep{MEZA2009, VANZELE2011, Monte2014, Wang2018}, also for reliable high frequency (HF) communications along short-, medium- and long-range paths, and for satellite communication, navigation, and positioning systems \citep{Gao2006, Le2009}. Therefore, continuous monitoring of the spatial and temporal variations of ionized plasma parameters, such as the F$_2$ layer peak electron density, N$_m$F$_2$, and TEC, are of particular interest. For instance, operators of space telecommunications need to know whether the ionospheric parameters indicate normal, quiet conditions in the ionosphere and plasmasphere or short-term perturbations of the ionospheric plasma, related to disturbances on the Sun and to geomagnetic processes.

Among a variety of techniques applied to probe the ionosphere, the GNSS is one of the most recognized sources of information. 
It provides instantaneous propagation delay, or equivalently, the Total Electron Content (TEC), allowing the estimation of GNSS derived TEC values for ground based reference stations.
 
3

The International GNSS Service \citep[IGS,][]{Johnston2017} and its associated Analysis Centres (ACs) have been providing global ionospheric maps (GIMs) without interruption since 1998. This GIMs exibit an unprecedented combination of accuracy, temporal and spatial resolution, and availability \citep{Mannucci1998, Schaer1999, Hernandez-Pajares2009}. 
The production of VTEC maps is employed in forecasting, nowcasting and characterization of space weather events \citep{Moulin2013}. Several research centers around the world are focused on the generation of near-real-time GNSS-derived TEC maps. These maps, in combination with other parameters, are very useful for alerting of geomagnetic disturbances. In Latin America, specially in Argentina, Brazil, Mexico and Peru, interest in space weather studies has become more important in the last decade \citep{Denardini2016a,Denardini2016b,Denardini2016c,Hysell2018, Valladares2012}. Currently, in Brazil and Mexico, regional GNSS-based TEC maps are systematically generated in order to assimilate them into space weather forecasting models \citep{Gonzalez2016, Takahashi2016}.

VTEC maps are usually combined in order to obtain ionospheric indexes, which in turn are designed to provide information about variations in the ionosphere-plasmasphere. In particular, the W-index is defined as the  logarithm of the ratio of the current value of VTEC, at a particular location, to the quiet background values VTEC$_\textrm{med}$ \citep{gulyaeva2014, gulyaeva2013}.

The objective of this work is to assert the quality of our regional ionospheric VTEC maps, generated in near-real-time, over Central and South America. Consequently, intercomparisons are made with openly-accessible reference VTEC maps of proven quality. Finally, a preliminary result on the computation of ionospheric W-index, during quiet and disturbed geomagnetic states, is made. The variations of this index, both in space and time, show important information about the state of the ionosphere under different scenarios.  

The paper is organized as follows. Section 2 describes the methodology used to compute the VTEC maps: the data cleaning stage (2.1), the procedure for the hardware delay calibration (2.2) and for TEC mapping in a near-real-time (2.3). The results about the quality assessment of our VTEC maps, and the subsequent computation of the ionospheric W-index maps, are discussed in section 3. Final remarks are given in section 4.

\section{Methodology}

\subsection{Data Preprocessing}
Before any computation a per station preprocessing and data cleaning is performed. This includes the application
of an appropriate time window, of an elevation cut off angle, of carrier phase wind-up corrections
and the determination of phase-continuous intervals (i.e., with constant ambiguity). The data cleaning is non-parametric
and consists in three steps.
Firstly, pairs of satellite-receiver phase and code links, in two bands, are optimally selected according to
the amount of available observations and the tracking modes, or channel attribute, employing the same default
priorities as the ones defined by \citet{gfzrnx}. The corresponding, undifferenced,
Melbourne-Wübbena (MW) linear combinations are screened for outliers and cycle slips (Table~\ref{wavelengths}).
Any possible receiver's clock inconsistency between phase and code observations is handled as a cycle slip.
Secondly, a per band, time-differenced, phase screening is performed, for unnoticed outliers and
cycle slips. These screenings, performed both for- and backward in time,
are repeated until no additional cycle slips, or outliers, are found. Furthermore, no attempt is made to correct any cycle slip.
Finally, MW and (cuasi) ionosphere-free (IF) linear combinations, within each phase-continuous interval, are
formed and modeled with low degree polynomials. Intervals resulting in residuals with root mean squared (RMS) greater
than given thresholds are rejected. In total, less than 7~\% of the original observations are generally left out, including those
observations bellow the elevation cut off angle.

Thereafter, a clean and single set of undifferenced carrier phase $\phi_{ij,k}$ (in cycles),
code pseudorange $C_{ij,k}$ (in meters) and signal-to-noise ratio $\textrm{SN}_{ij,k}$ observables,
between each pair of satellite $i$ and receiver $j$, and for each tracked band $k$,
is obtained. This data preprocessing is performed with the \mbox{Fortran 2008} + OpenMP, in-house developed, software \emph{AGEO}
 (library for Geodetic and Orbital Analysis or \emph{biblioteca de Análisis GEodésico y Orbital}, in Spanish).
In addition, it is externally parallelized, on a per station basis,
by means of the \emph{GNU parallel} software tool \citep{parallel}.

\begin{table}[ht]
\caption{Pairs of bands employed in the Melbourne-Wübbena (MW) data screening, with their corresponding
wide-lane wavelengths $\lambda_{\textrm{WL}}$ (in meters). Those pairs of bands employed in the computation of geometry-free (GF) linear combinations
and inter-frequency biases (IFBs) are also indicated.
Observation codes according to RINEX version 3.03 \citep{rinex}.}
\label{wavelengths}
\centering
\begin{tabular}{l c c c c}
\hline
 GNSS  & \multicolumn{2}{c}{Bands} & $\lambda_{\textrm{WL}}$ & GF \& IFB\\
\hline
GPS & L2 & L5 & 5.86 & no \\
 & L1 & L2 & 0.86 & yes \\
 & L1 & L5 & 0.75 & yes \\
GLONASS & L2 & L3 & 6.82 & no\\
 & L1 & L2 & 0.84 & yes \\
 & L1 & L3 & 0.75 & yes \\
Galileo & L7 & L5 & 9.77 & no\\
 & L7 & L6 & 4.19 & no\\
 & L6 & L8 & 3.54 & no\\
 & L6 & L5 & 2.93 & no\\
 & L1 & L6 & 1.01 & yes \\
 & L1 & L7 & 0.81 & yes \\
 & L1 & L8 & 0.78 & yes \\
 & L1 & L5 & 0.75 & yes \\
BeiDou & L6 & L7 & 4.88 & no\\
 & L2 & L6 & 1.04 & yes \\
 & L2 & L7 & 0.84 & yes \\
\hline
\end{tabular}
\end{table}

\subsection{Hardware Delays Calibration}\label{a_priori}
Once per hour inter-frequency biases (IFBs) are estimated from carrier-to-code
leveled geometry-free (GF) linear combinations \citep[see, e.g.,][]{Spits2012}.
These hardware delays are solved simultaneously with spherical harmonic (SH) coefficients of a single-layer VTEC
representation \citep{Schaer1999},
assuming that all free electrons are constrained to an infinitesimally thin layer at a height of \mbox{450 km}.
Here only independent linear combinations are employed, between
selected pairs of bands (Table~\ref{wavelengths}). Hence, only independent IFBs are computed.
Also, no closing restriction is imposed, resulting in
\mbox{satellite-receiver-,} pair-of-bands-specific IFBs.  Moreover, different tracking
modes or channel attributes, between each pair of satellite-receiver links, are taken into account. 
The estimation is made by means of a weighted least squares adjustment, performed also with the \emph{AGEO} software, and executed in one single step,
involving the most recent observations available, from all ground
stations and all satellites, within the previews 24 hours (i.e., a 24 hours rolling- or moving-window).

In practice, after preprocessing the raw observations corresponding to each pair satellite $i$ and receiver $j$, and
for each pair of bands $k$ and $l$, all possible carrier-to-code leveled GF linear
combinations $\widetilde{L}_{\textrm{GF},ij,kl}$ (in meters) are computed by
\begin{equation}
\widetilde{L}_{\textrm{GF},ij,kl} = L_{\textrm{GF},ij,kl} - \langle L_{\textrm{GF},ij,kl} - C_{\textrm{GF},ij,kl} \rangle 
\end{equation}
where $L_{\textrm{GF},ij,kl} = \lambda_k \phi_{ij,k} - \lambda_l  \phi_{ij,l}$ are the non-leveled
GF linear combinations in phase (in meters) and $C_{\textrm{GF},ij,kl} = C_{ij,l} -  C_{ij,k}$ are the 
corresponding linear combinations in code pseudorange (in meters), being $\lambda_k$ and $\lambda_l$
the wavelengths of each band (in meters). Here the average is computed within each phase-continuous interval,
under the assumption of stable hardware delays.
This commonly used methodology reduces the observations noise, from code to phase levels, and avoids the estimation of phase ambiguities,
but it could also introduce some systematic errors \citep[see, e.g.,][]{Ciraolo2007,Spits2012}.
In addition, each GF observation is weighted
according to three factors: the instantaneous satellite elevation, the amount of observations employed during the carrier-to-code leveling
(i.e., the length of each phase-continuous interval) and the corresponding navigational system \citep[GPS, GLONASS, Galileo or BeiDou, see][]{Ren2016}.
Then, the full set of GF observations is represented as
\begin{equation}
\widetilde{L}_{\textrm{GF},ij,kl} = \alpha_{kl}\operatorname{MF}(z)\operatorname{VTEC}(\mu,t)+\textrm{IFB}_{ij,kl}
\end{equation}
where $\operatorname{MF}(z)$ is the Modified Single Layer Model (MSLM) mapping function \citep{Schaer1999}, being $z$ the zenith distance
of satellite $i$ as seen by receiver $j$ (in radians),
$\textrm{IFB}_{ij,kl}$ are the corresponding specific IFBs (in meters), $\alpha_{kl}$ is a proportionality constant (in meters per TECU,
where 1 TECU is equivalent to $10^{16}$ free e$^-$ per squared meter)
\begin{equation}
\alpha_{kl} = 40.3\times10^{16}\left(\frac{1}{f_k^2} - \frac{1}{f_l^2}\right)
\end{equation}
being $f_k$ and $f_l$ the frequencies of each band (in Hertz), whereas the VTEC is expressed as a
SH expansion in a sun-fixed frame
\begin{equation}
\operatorname{VTEC}(\mu,t) = \sum_{n=0}^{n_\textrm{max}}\sum_{m=0}^{n} \overline{P}_{nm}(\sin\mu)
\big(a_{nm} \cos(m\,t) + b_{nm} \sin(m\,t)\big).
\end{equation}
Here $a_{nm}$ and $b_{nm}$ are the coefficients of the SH expansion (in TECU), with maximum degree $n_{max}$,
whereas $\overline{P}_{nm}$ are the corresponding
Real Associated Legendre Functions \citep[$4\pi$ normalized, see for example][]{Wieczorek}, $t$
is the Local Time (LT, in radians) and $\mu$ is the modified dip latitude (also in radians). In this case
the algorithms issued by the \citet{galileo}, together with the corresponding global grid, are
employed for the computation of $\mu$.

As only regional observations are employed, the 24 hours time window helps into decoupling the hardware delays, from
the ionospheric parameters, by always including in the adjustment observations spanning 24 hours of LT. Furthermore,
a single set of constant coefficients $a_{nm}$ and $b_{nm}$, loosely constrained to a \emph{zero} ionosphere,
is estimated for the entire time span, resulting in a mean (daily) VTEC representation.
For the same reason a SH expansion of
low degree is employed, in order to avoid ill conditioned normal equations \citep{haines1985}, which in turn could produce
mapping artifacts,
particularly at the boundaries of the region. On the other hand,
the $\textrm{IFB}_{ij,kl}$ are also parametrized as constants, and estimates with mean observational epoch at the middle of each moving-window
are obtained. These hardware delays are also loosely constrained to their most recently estimated values.
In fact, the main result of this hourly adjustment are precisely these decoupled $\textrm{IFB}_{ij,kl}$ estimates.
%However, they result systematically half a  day old. 

\subsection{Near-Real-Time TEC Mapping}
Every 15 minutes, and also by means of a weighted least squares adjustment, both the IFBs and the SH
coefficients for the regional VTEC representation are updated.
In essence, the same software, methodology and parametrization described in the previews section are employed.
However, in this case only the most recent observations available, within a one hour moving-window,
are used. In addition, here four sets of pice-wise constant SH coefficients are estimated, each one valid
for a quarter of an hour.  Furthermore, the $\textrm{IFB}_{ij,kl}$ parameters are now actively constrained to their
most recent, hourly, and decoupled estimates. In practice, this results in new hardware delays estimates
that are simultaneously up-to-date (i.e., less than 30 minutes old) and decoupled from the coefficients of the SH expansion. Thereafter, tracks
of instantaneous, VTEC estimates are obtained from the original GF observations by
\begin{equation}
{\operatorname{VTEC}}_{ij,kl,\varphi\lambda} =  \big(\alpha_{kl}\operatorname{MF}(z)\big)^{-1}\big(\widetilde{L}_{\textrm{GF},ij,kl} - {\operatorname{IFB}}_{ij,kl}\big)
\end{equation}
where $\varphi$ and $\lambda$ are the geographic latitude and longitude, respectively, of the ionospheric pierce points (IPPs), that is, the
intersection point between the instantaneous satellite-receiver line-of-sight with the single layer of the model.
Similarly, traces of slant TEC (STEC) estimates can be computed by
\begin{equation}
{\operatorname{STEC}}_{ij,kl,\varphi\lambda} =  \alpha_{kl}^{-1}\big(\widetilde{L}_{\textrm{GF},ij,kl} - {\operatorname{IFB}}_{ij,kl}\big).
\end{equation}
At this point two representations of the current state of the regional ionospheric TEC are available. In one hand,
an analytical representation, given by the coefficients of a low degree SH expansion in $\mu$ and $t$,
with mean epoch at the middle of the latest 15 minutes of the observational window. On the other hand, a discreet
and huge set of instantaneous $\textrm{VTEC}_{ij,kl,\varphi\lambda}$ estimates, along the IPP tracks, during the same interval.
In fact, the issued TEC product is obtained by mapping these tracks in the space domain.

This postprocessing of the $\textrm{VTEC}_{ij,kl,\varphi\lambda}$ estimates comprises three steps, all 
performed with the \emph{Generic Mapping Tools} software package \citep[GMT,][]{gmt}. Firstly, all available estimates
are averaged within the cells of a uniform $0.5 \times 0.5$ degrees grid, previously discarding cells with very few observations,
and effectively resulting in $N$ space- and time-averaged $\langle\textrm{VTEC}\rangle_p$ values, for $p=1,\ldots,N$.
Secondly, this regular grid is approximated, using a generalized Green's function for continuous curvature spherical spline in tension
\citep{greenspline}, by
\begin{equation}
\operatorname{VTEC}(\varphi,\lambda) = c_0 + \sum_{p=1}^M c_p\,g(\varphi,\lambda,\varphi_p,\lambda_p)
\end{equation}
where $\varphi$ and $\lambda$ are arbitrary coordinates, $M\le N$ is the number of employed coefficients,
$\varphi_p$ and $\lambda_p$ are the coordinates of the corresponding cells, $c_0$
is the mean VTEC over all populated cells (in TECU), $g$ is the generalized Green's function and $c_p$
are the spline coefficients (also in TECU), solved for by
Singular-Value Decomposition (SVD) on the square linear system
\begin{equation}
\langle\textrm{VTEC}\rangle_p  - c_0 = \sum_{q=1}^N c_q\,g(\varphi_p,\lambda_p,\varphi_q,\lambda_q)
\end{equation}
and retaining only those $M$ eigenvalues whose ratios, to the largest,
are greater than a given threshold. While we empirically determined optimal (fixed) values for both the tension and the threshold,
searching over thousand of maps for minimization of the misfits, the number $M$ of contributing eigenvalues is dynamically
determined, every time, to accommodate the variance of the current data. 
That is, the more spatial variability in the regional VTEC the more eingenvalues are retained
in the mapping procedure.
Finally, the adjusted function is evaluated on a uniform $0.5 \times 0.5$ degrees grid and areas far away from
IPP tracks are automatically masked out. The resulting grid constitutes
the actual, near-real-time, regional TEC map produced by the system,
as no additional
postprocessing (e.g., smoothing) is required nor performed.

\section{Results}

\subsection{Year-round Intercomparison with Global Products}
To evaluate the quality of the produced maps, and particularly the possible presence of systematic biases,
we compared them with several IGS final VTEC products,
provided in IONEX format, and
computed by several IGS Ionosphere Associated Analysis Centers (IAACs): Center for Orbit Determination in Europe \citep[CODE, Switzerland; see][]{Schaer1999},
European Space Agency/European Space Operations Centre \citep[ESA/ESOC, Germany, see][]{Feltens2007},
IGS \citep[see][]{Hernandez-Pajares2009}, Jet Propulsion Laboratory/National
Aeronautics and Space Administration \citep[JPL/NASA, USA; see][]{Mannucci1998} and
Universitat Politècnica de Catalunya \citep[UPC, Spain; see][]{Hernandez-Pajares1999,Orus2005}.
From UPC we employed both, their standard and their high rate products.
We also included in the analysis VTEC products from two additional IGS ACs: Natural Resources Canada \citep[NRCAN, Canada; see][]{Ghoddousi-Fard2011}
and Wuhan University \citep[WHU, China; see][]{Wang2018}.
These products are usually available with latencies of a few days or, at best, several hours.
In addition, we also included in the intercomparison the (non-IGS) global and high resolution
TEC products provided by the Massachusetts
Institute of Technology \citep[MIT Haystack Observatory, USA; see][]{Rideout2006}.

The comparison extends a full year, from June 1, 2017 to May 31, 2018,
and it was performed on a map by map basis (i.e., epoch by epoch).
In order to assess the expected differences we performed
the same one-to-one comparison between pairs of global products. Although these maps
have global coverage, the comparison was restricted to the area covered by our regional maps, that is, between
$80^\circ$~S and $40^\circ$~N in latitude and $110^\circ$~W and $0^\circ$~E in longitude
\citep[see Figure~1 in][]{mendoza2019a}.
Also, no spatial or temporal interpolation was performed. Rather, only VTEC samples at common epochs, and exactly
the same reported locations, were differenced. For this reason, and before the comparisons, our high resolution maps
were downsampled. Thus, the results were controlled by the standard \mbox{$5\times 2.5$} degrees
spatial sampling (in longitude and latitude, respectively) of the IGS products or,
alternatively, by the \mbox{$1\times 1$} degrees spatial sampling of the MIT products.

For this analysis, instead of the real-time data streams, we employed daily observational and navigational
RINEX files available at the servers of the respective data providers. However, to reproduce exactly the
results of the near-real-time system, we only used data from those GNSS stations that are actually
accessible in real-time, leaving all off-line stations out of the analysis. We also employed the very same
broadcasted orbits and satellite clocks, and no other products. In addition, we
followed exactly the same two-steps methodology previously described. That is, a first step resulted in
IFBs estimates, from a 24 hours observational moving-window, while in a second and final step the
TEC maps were produced, from a 15 minutes moving-window. To speed up this year-round analysis,
and although some of the selected products are currently provided at a higher rate (e.g.,by
CODE, NRCAN and particularly UPC and MIT), we computed maps with 2 hours of temporal sampling, following
the classical IGS standard practice.

\begin{figure}[ht]
\centering
\includegraphics[width=\textwidth]{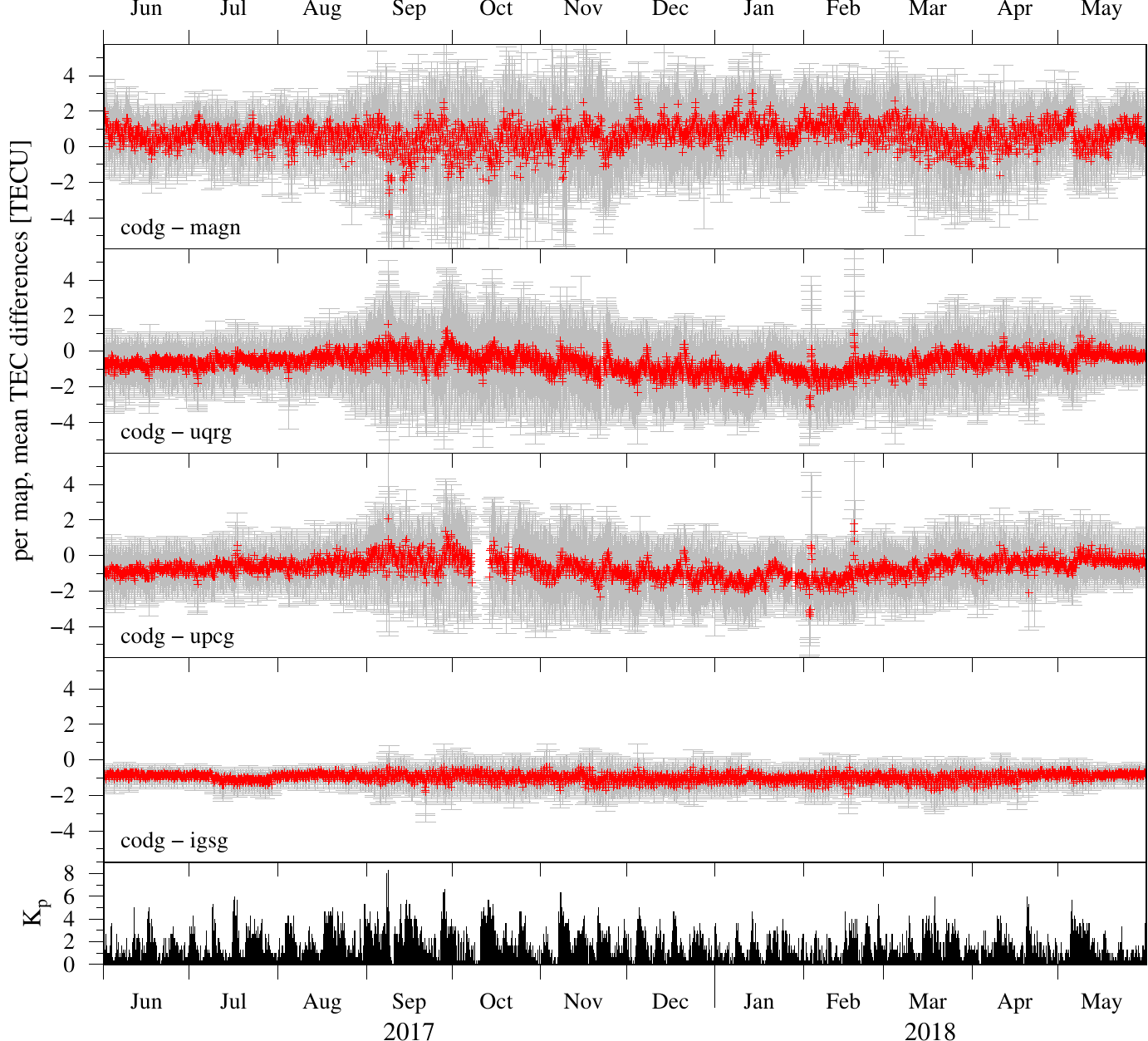}
\caption{Examples of mean differences and standard deviations, per map, resulting from the year-round one-to-one
VTEC products comparisons: between a final IGS
and our near-real-time product (codg and magn, respectively), between two final IGS products
(codg and the high rate uqrg), between two final IGS products (codg and upcg, noting
that no upcg IONEX files were available for October 8--13 and 21, 2017 and for January 28, 2018) and
between a final and the combined IGS product (codg and igsg, respectively).
The global $K_p$ index, provided by the GeoForschungsZentrum (GFZ, Germany), is also plotted.}
\label{timeseries}
\end{figure}

The year-round (and regional) comparison shows the existence of systematic differences
 between all the analyzed pairs of VTEC products (Table~\ref{oneyear}). In average, our near-real-time VTEC maps show a very
good agreement with the maps produced by ESA/ESOC, CODE, UPC (high rate) and especially NRCAN, resulting for all the cases
in a mean bias lower or equivalent to 1 TECU, comparable with the differences found between pairs of global products.
Nevertheless, while there seems to be no significantly biases between both,
our products and the ones from NRCAN and between them and the ones from CODE, a small systematic bias do exists
between the former and our products. The reason for this seemingly discordant results is simple: given their global coverage,
the comparisons between pairs of IGS products span the entire area mapped \citep[Figure~1 in][]{mendoza2019a}, whereas
those comparisons involving our regional product are mostly restricted to the land, leaving large portions of the oceans
out of the analysis, and this is evident in the lower number of common TEC samples found (Table~\ref{oneyear}). 
This contributes also to the higher mean standard deviation encountered while comparing our maps with
the other products, both in average and individually (Figure~\ref{timeseries}). Indeed,
not only a smaller number of differences are averaged, also the smoothest areas of the IGS maps
over the oceans, where no actual GNSS observations were available, are systematically left out of these comparisons.
At the same time, all comparisons show, to a greater or lesser extent, smaller variance
during the southern winter (i.e., June, July and August). This is probably due to the lower, regional, mean ionospheric TEC
in that season.

\begin{table}
\caption{Year-round one-to-one comparison between selected (GNSS-based) VTEC products, from June 1, 2017 to May 31, 2018:
codg (CODE), emrg (NRCAN), esag (ESA/ESOC),
igsg (IGS, combination of codg and jplg), jplg (JPL/NASA), mapgps (MIT),
upcg (UPC), uqrg (UPC, high rate), whug (WHU) and magn (our near-real-time product).
The mean difference $\overline{x}_{A-B}$ and
mean standard deviation $\overline{\sigma}_{A-B}$, over all compared maps, are expressed in TECU.}
\label{oneyear}
\centering
\begin{small}
\begin{spacing}{0.65} % otherwise this table is to long for review mode, comment for final version!!!
\begin{tabular}{c c r r c c}
\hline
\multicolumn{2}{c}{Products} & & & TEC & TEC\\
 A & B &  $\overline{x}_{A-B}$ & $\overline{\sigma}_{A-B}$ & Maps & Samples\\
\hline
codg & magn & $    0.7 $ & $    1.9 $ &   4380 &   1965432 \\
 & emrg & $    0.1 $ & $    2.1 $ &   8758 &   9869733 \\
 & esag & $   -0.1 $ & $    1.6 $ &   4368 &   4922736 \\
 & igsg & $   -1.0 $ & $    0.6 $ &   4380 &   4936260 \\
 & jplg & $   -2.2 $ & $    1.3 $ &   4380 &   4936260 \\
 & mapgps & $ 2.3 $ & $ 2.3 $ & 8568 & 1232740 \\
 & upcg & $   -0.7 $ & $    1.5 $ &   4284 &   4828068 \\
 & uqrg & $   -0.7 $ & $    1.7 $ &   8736 &   9845472 \\
 & whug & $    1.7 $ & $    4.4 $ &   5280 &   5950560 \\
emrg & magn & $    0.1 $ & $    2.4 $ &   4380 &   1965367 \\
 & esag & $   -0.2 $ & $    2.6 $ &   4368 &   4922471 \\
 & igsg & $   -1.0 $ & $    2.1 $ &   4380 &   4935994 \\
 & jplg & $   -2.2 $ & $    2.4 $ &   4380 &   4935994 \\
 & mapgps & $    2.1 $ & $ 2.2 $ & 8566 & 1232401 \\
 & upcg & $   -0.8 $ & $    2.0 $ &   4284 &   4827806 \\
 & uqrg & $   -0.7 $ & $    2.3 $ &   8734 &   9842685 \\
 & whug & $    1.7 $ & $    4.6 $ &   5278 &   5947974 \\
esag & magn & $    0.9 $ & $    2.2 $ &   4368 &   1960492 \\
 & igsg & $   -0.8 $ & $    1.7 $ &   4368 &   4922736 \\
 & jplg & $   -2.0 $ & $    2.0 $ &   4368 &   4922736 \\
 & mapgps & $ 2.4 $ & $ 2.8 $ & 4272 & 614605 \\
 & upcg & $   -0.5 $ & $    1.7 $ &   4272 &   4814544 \\
 & uqrg & $   -0.5 $ & $    2.3 $ &   4356 &   4909212 \\
 & whug & $    2.4 $ & $    5.0 $ &   4368 &   4922736 \\
igsg & magn & $    1.6 $ & $    1.9 $ &   4380 &   1965432 \\
 & jplg & $   -1.2 $ & $    0.7 $ &   4380 &   4936260 \\
 & mapgps & $ 3.3 $ & $ 2.3 $ & 4284 & 616229 \\
 & upcg & $    0.3 $ & $    1.5 $ &   4284 &   4828068 \\
 & uqrg & $    0.3 $ & $    1.6 $ &   4368 &   4922736 \\
 & whug & $    3.2 $ & $    5.1 $ &   4380 &   4936260 \\
jplg & magn & $    2.9 $ & $    2.1 $ &   4380 &   1965432 \\
 & mapgps & $ 4.6 $ & $ 2.4 $ & 4284 & 616229 \\
 & upcg & $    1.5 $ & $    1.8 $ &   4284 &   4828068 \\
 & uqrg & $    1.5 $ & $    1.8 $ &   4368 &   4922736 \\
 & whug & $    4.4 $ & $    5.3 $ &   4380 &   4936260 \\
mapgps & magn & $ -2.0 $ & $ 2.8 $ &  4284 & 10182540 \\
 & upcg & $ -2.7 $ & $ 2.3 $ & 4188 & 603294 \\
 & uqrg & $ -2.7 $ & $ 2.2 $ & 33470 & 4816557 \\
 & whug & $ -0.1 $ & $ 4.2 $ & 5160 & 749695 \\
upcg & magn & $    1.1 $ & $    1.9 $ &   4128 &   1848581 \\
 & uqrg & $    0.0 $ & $    1.1 $ &   4272 &   4814544 \\
 & whug & $    2.9 $ & $    4.9 $ &   4284 &   4828068 \\
uqrg & magn & $    1.0 $ & $    1.9 $ &   4369 &   1960667 \\
 & whug & $    2.3 $ & $    4.7 $ &   5268 &   5937036 \\
whug & magn & $   -2.7 $ & $    5.0 $ &   4377 &   1963993 \\
\hline
\end{tabular}
\end{spacing}
\end{small}
\end{table}

\subsection{Differential STEC Evaluation}

In order to independently assess the accuracy of the produced VTEC maps we applied a differential STEC (dSTEC)
test developed by the IGS Ionosphere Working Group (IIWG) for the evaluation,
and relative weighting,
of their Global Ionosphere Maps (GIMs) \citep[see, for example,][]{Orus2005,Orus2007,Roma-Dollase2018}. In essence,
the test is based on the ability to make highly accurate dSTEC measurements, on the order
of $10^{-2}$ TECU \citep{HernandezPajares2017,Coster2013},
and to compare them with synthetic (i.e., mapped) dSTEC values. In fact, we employed
the very same implementation of the test as described in detail by \citet{HernandezPajares2017},
the only difference being our extension of the test to the multi-frequency case.

The analysis is performed on a per station basis, involving only GNSS stations that were
not employed for the computation of the VTEC maps being evaluated. Firstly, and after preprocessing the corresponding
raw data (e.g., outliers rejection, phase wind-up correction, etc.), observed $\textrm{dSTEC}_o$ (in TECU)
are obtained from (non-leveled) carrier phase GF linear combinations (in meters) by
\begin{equation}
\textrm{dSTEC}_o(t_s) = \alpha_{kl}^{-1}\big(L_{\textrm{GF},ij,kl}(t_s) - L_{\textrm{GF},ij,kl}(t_r)\big) \quad \textrm{with } t_r\ne t_s,
\end{equation}
taking advantage of the total cancellation of the phase ambiguities within each phase-continuous interval.
Here $t_r$ (in hours) represents a reference epoch, when the satellite reaches its
minimum zenith distance within each phase-continuous interval, whereas $t_s$ (in hours) are all other sample epochs
within the same phase interval. Here we employed a sampling rate of 60 seconds and, following the convention stated
by \citet{HernandezPajares2017}, only GF observations no more than 900 seconds apart were differenced. This
results in a maximum of 30 $\textrm{dSTEC}_o$ samples per phase-continuous interval, regardless of its total length.
In addition, the corresponding
zenith distances $z_r$ and $z_s$ (both in radians), with $z_r \ne z_s$, are stored for subsequent use.
Secondly, and for each observed $\textrm{dSTEC}_o$ sample, synthetic $\textrm{dSTEC}_m$ values (in TECU) are computed by
\begin{equation}
\textrm{dSTEC}_m(t_s) = \operatorname{MF}(z_s)\operatorname{VTEC}(\varphi_s,\lambda_s,t_s) -
 \operatorname{MF}(z_r)\operatorname{VTEC}(\varphi_r,\lambda_r,t_r)
\end{equation}
where $\varphi_r,\lambda_r$ and $\varphi_s,\lambda_s$ (in degrees) are the coordinates of
the corresponding IPPs. Here both $\operatorname{VTEC}(\varphi,\lambda,t)$ are obtained, following \citet{ionex}, by 
temporal interpolation between consecutive \emph{rotated} TEC maps
\begin{equation}
\operatorname{VTEC}(\varphi,\lambda,t) = \frac{T_{i+1} - t}{T_{i + 1}-T_i}\operatorname{VTEC}_i(\varphi,\lambda_i')+
\frac{t - T_i}{T_{i + 1}-T_i}\operatorname{VTEC}_{i+1}(\varphi,\lambda_{i+1}')
\end{equation}
being $T_i$ and $T_{i+1}$ the epochs of the corresponding maps (in hours), with $T_i<t<T_{i+1}$, whereas the
rotated longitudes $\lambda_i'=\lambda+15\,(t-T_i)$ and $\lambda_{i+1}'=\lambda+15\,(t-T_{i+1})$ compensate
the strong correlation between the ionospheric TEC and the (longitude of the) subsolar point. Within
each map, the $\textrm{VTEC}_i$ and $\textrm{VTEC}_{i+1}$ are spatially interpolated by a simple
4-point bilinear algorithm \citep[see also][]{ionex}. Finally, the observed minus computed $\Delta \textrm{dSTEC}$
(in TECU) are obtained
\begin{equation}
\Delta \operatorname{dSTEC}(t_s) = \operatorname{dSTEC}_o(t_s) - \operatorname{dSTEC}_m(t_s).
\end{equation}
In turn, the RMS of $\Delta \textrm{dSTEC}$, per station, can be computed.

\begin{figure}[ht]
\centering
\includegraphics[scale=1]{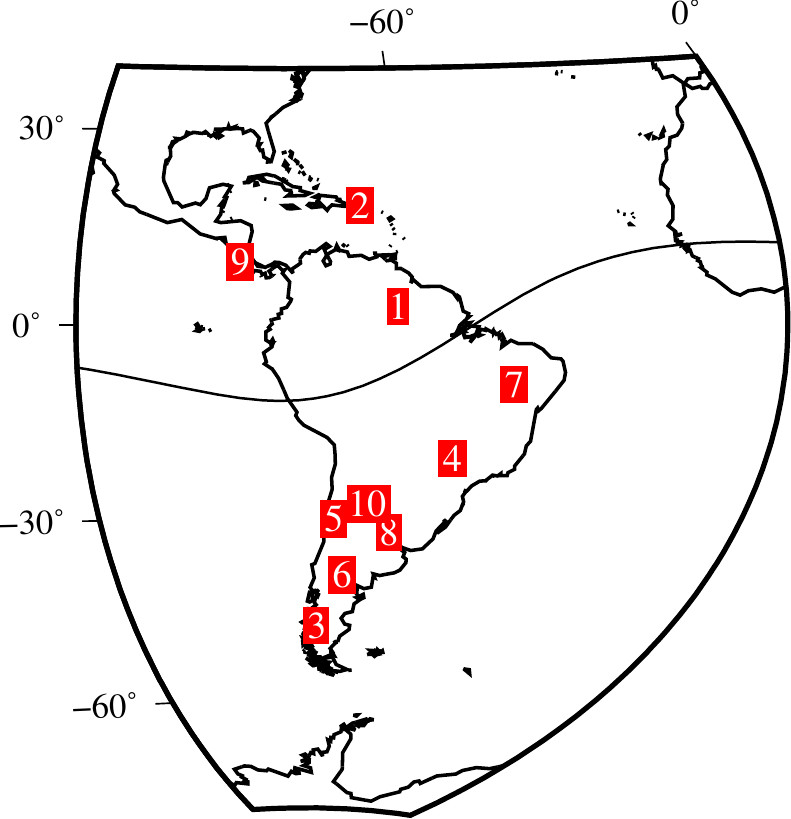}
\caption{Location of the off-line GNSS stations, listed in Table~\ref{dstec}, and employed for the dSTEC evaluation of the
near-real-time TEC maps. For convenience the geomagnetic equator is also plotted.}
\label{dstecmap}
\end{figure}

In practice, we employed daily RINEX files from ten GNSS stations distributed over the study area (Figure~\ref{dstecmap}).
In addition to files from the mentioned data providers we also employed observations from off-line GNSS stations
supplied by the Centro Sismológico Nacional (CSN, Chile). The analysis was repeated in four independent
days, during the years 2017 and 2018, near the ascending equinox, the descending equinox, the summer solstice
and the winter solstice. Also, the TEC maps employed in this analysis are the very same produced for
the year-round comparison with the IGS GIMs. However, for these four particular days, additional maps were produced
in order to achieve the standard 15 minutes sampling rate of the monitoring system.

In summary, the observed dSTECs are fairly reproduced by the synthetic values,
implying that the near-real-time maps, in combination with the corresponding mapping function, are
capable of representing the regional ionospheric VTEC with an average
accuracy better than \mbox{1 TECU} (Table~\ref{dstec}).
Finally, the three stations leading to a total RMS $> 0.7$ TECU
are located in areas where the IPPs coverage is, systematically, not optimal
\citep[especially near BOAV and PUMO, but to a lesser extent also near PISR, see Figure~1 in][right]{mendoza2019a}.
This suggests that the monitoring system could benefit from the use of
additional data from GNSS stations located in these specific areas.

\begin{table}
\caption{Per station daily and total RMS (in TECU) resulting from the dSTEC evaluation of the
near-real-time ionospheric VTEC maps.}
\label{dstec}
\centering
\begin{tabular}{c c c c c c c c c}
\hline
\multicolumn{2}{c}{GNSS Station} & \multicolumn{3}{c}{2017} & 2018 & Total & $\Delta\textrm{dSTEC}$\\
\# & ID &  Jun 21 & Sep 22 & Dec 21 & Mar 20 & RMS & Samples\\ 
\hline
1 & BOAV & 0.59 & 1.17 & 1.01 & 0.75 & 0.91 & 38611\\
2 & BYSP & 0.44 & 0.43 & 0.32 & 0.50 & 0.43 & 24498\\
3 & CCHR & 0.47 & 0.65 & 0.84 & 0.54 & 0.61 & 16690\\
4 & ILHA & 0.28 & 0.64 & 0.89 & 0.58 & 0.66 & 35539\\
5 & JUNT & 0.84 & 0.41 & 0.42 & 0.73 & 0.64 & 12793\\
6 & MA01 & 0.26 & 0.52 & 0.86 & 0.46 & 0.59 & 23448\\
7 & PISR & 0.43 & 0.80 & 1.11 & 0.86 & 0.84 & 20104\\
8 & PRNA$^\dag$ & -- & 0.23& 0.62 & 0.42 & 0.50 & 17036\\
9 & PUMO & 0.51 & 0.73 & 0.51 & 1.44 & 0.89 & 23461\\
10 & TERO & 0.30 & 0.93 & 0.54 & 0.65 & 0.65 & 39588\\
\hline
\multicolumn{8}{l}{$^{\dag}$No daily RINEX file, for June 21, 2017, was available.}
\end{tabular}
\end{table}

\subsection{W-Index}
The spatial and temporal variations of ionized plasma parameters are of particular interest in many applications, such as radiocommunications, and space-based navigation and positioning. There are many kind of ionospheric indexes proposed to describe ionosphere-plasmasphere variations. In this work we focus on those indexes derived from GNSS measurements through VTEC estimates. In particular, the deviation from the quiet median which is defined as
\begin{equation}
\label{log}
\operatorname{DEV}(\textrm{VTEC}) = \log \left( \frac{\textrm{VTEC}\,(t)}{\textrm{VTEC}_\textrm{med}} \right)
\end{equation}
where $\textrm{VTEC}_\textrm{med}$ is the quiet reference 27-days-running median prior to the epoch of observation \citep{gulyaeva2014, gulyaeva2013}. Therefore, from $\operatorname{DEV}(\textrm{VTEC})$ the W-index is defined as a measure of the ionosphere-plasmasphere state, and it could be computed at each point of a regular grid (Table~\ref{w-index}).

\begin{table}
\caption{Categories of the ionospheric weather W-index corresponding to the logarithmic deviation from the median.}
\label{w-index}
\centering
\begin{center}
\begin{tabular}{cc} 
%\begin{tabular}{c c }
\hline
W-index & $\operatorname{DEV}(\textrm{VTEC})$ \\
\hline
$\phantom{-}4$ & $\textrm{DEV} > 0.301$ \\
$\phantom{-}3$ & $0.155 < \textrm{DEV} \leq 0.301$ \\
$\phantom{-}2$ & $0.046 < \textrm{DEV} \leq   0.155$ \\
$\phantom{-}1$ & $0 < \textrm{DEV} \leq   0.046$ \\
$\phantom{-}0$ & $ \textrm{DEV}  = 0 $ \\
 $-1$          & $-0.046 < \textrm{DEV} \leq   0$ \\
 $-2$          & $-0.155 <  \textrm{DEV} \leq   -0.046$ \\
 $-3$          & $-0.301 < \textrm{DEV} \leq   -0.155$ \\
 $-4$          & $ \textrm{DEV} < -0.301$ \\
\hline
\end{tabular}
\end{center}
\end{table}

\begin{figure*}[htbp]
\begin{center}
%\textbf{\hspace{5mm} Dst and Kp geomagnetic indices}\par\medskip
%\subfigure{\includegraphics[height=100pt, width=260pt]{./dst.jpg}}\vspace{.5mm}
%\subfigure{\includegraphics[height=100pt, width=260pt]{./kp.jpg}}\vspace{.5mm}
\includegraphics[width=\textwidth]{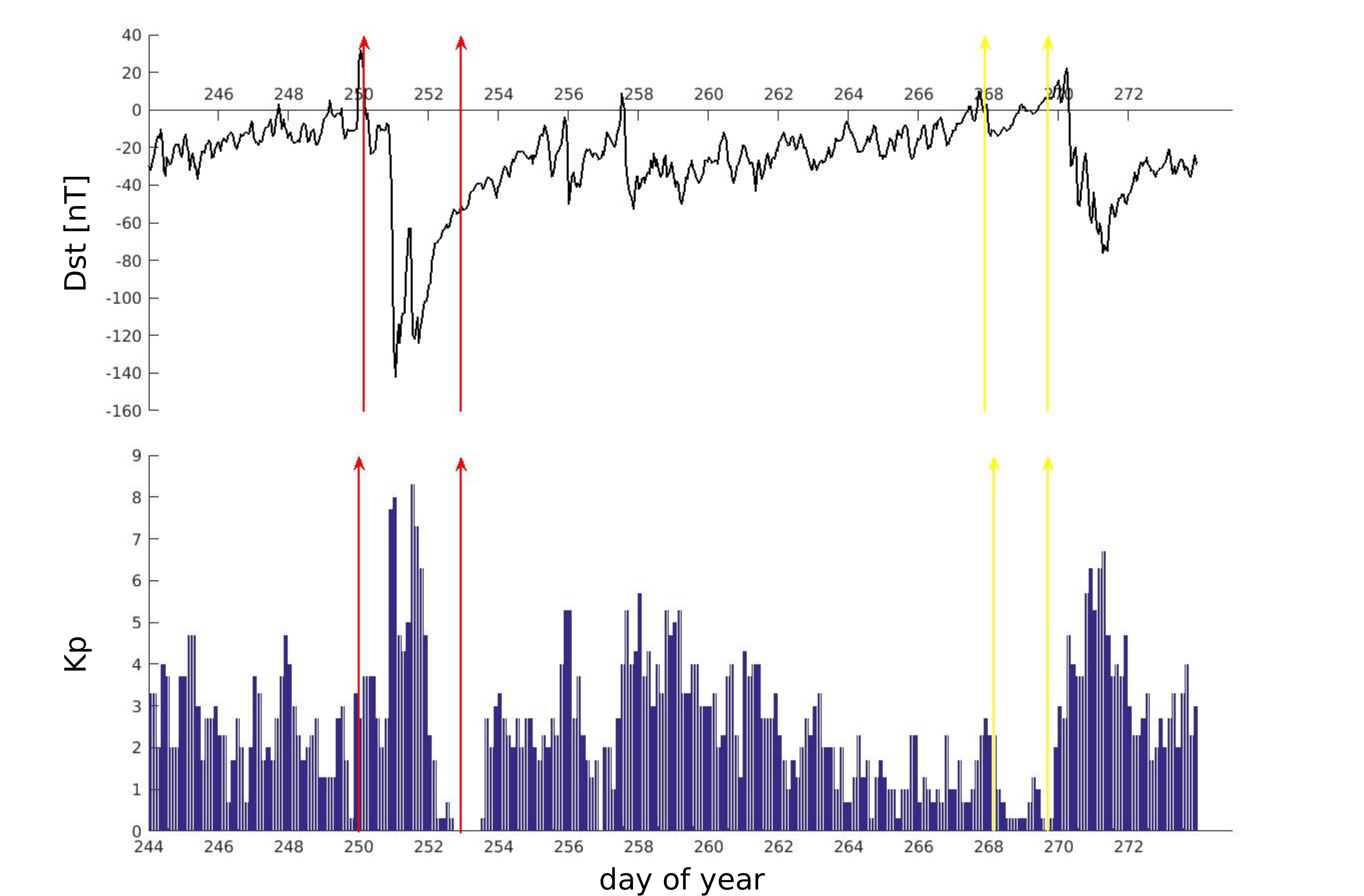}
\end{center}
\caption{Values of Dst and Kp indices for September 2017}
\label{Dst_Kp}
\end{figure*}

\begin{figure*}[htbp]
\begin{center}
\includegraphics[width=\textwidth]{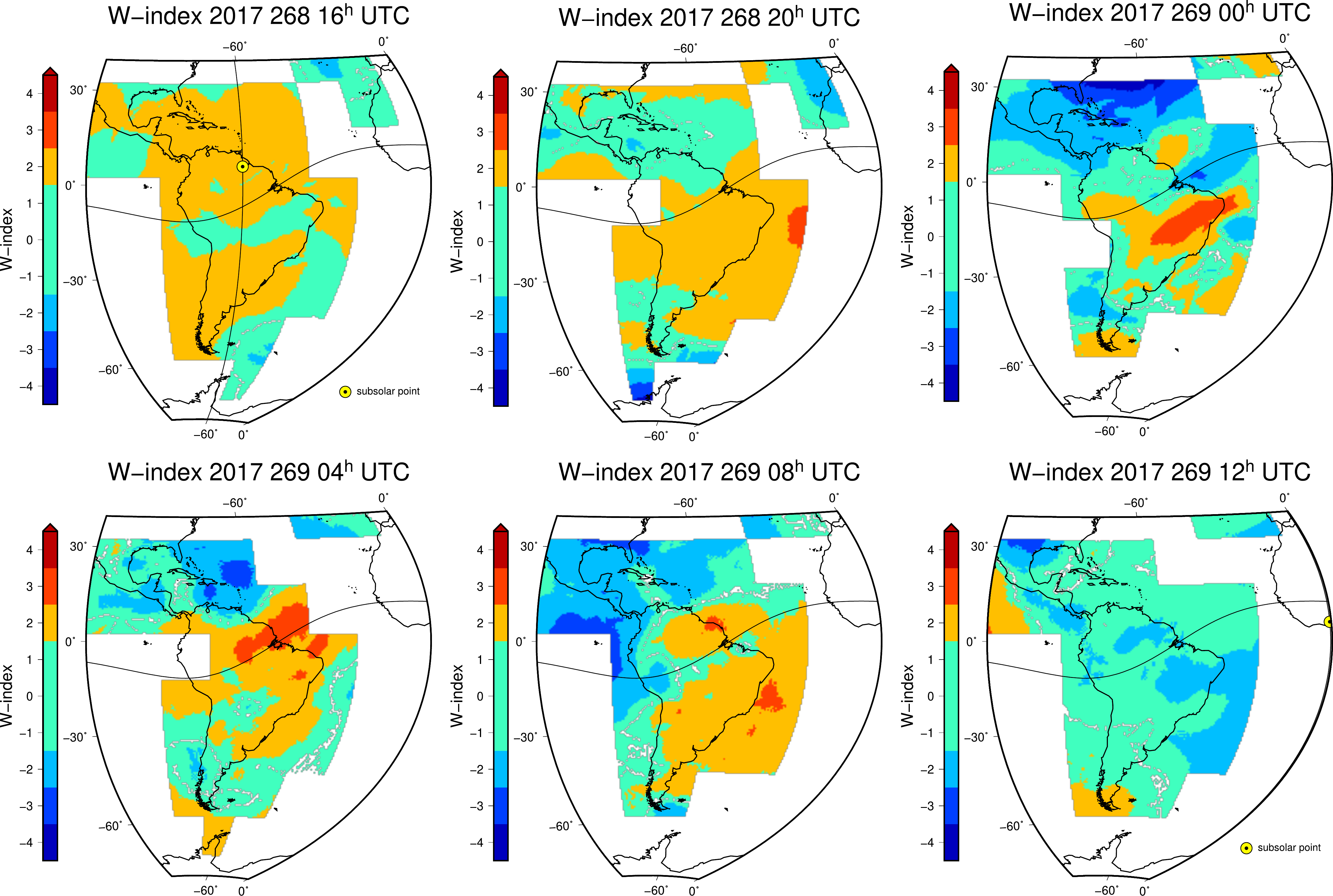}
\end{center}
\caption{Ionospheric W-index maps during quiet and moderate disturbance period}
\label{QQ_QD}
\end{figure*}

\begin{figure*}[htbp]
\begin{center}
\includegraphics[width=\textwidth]{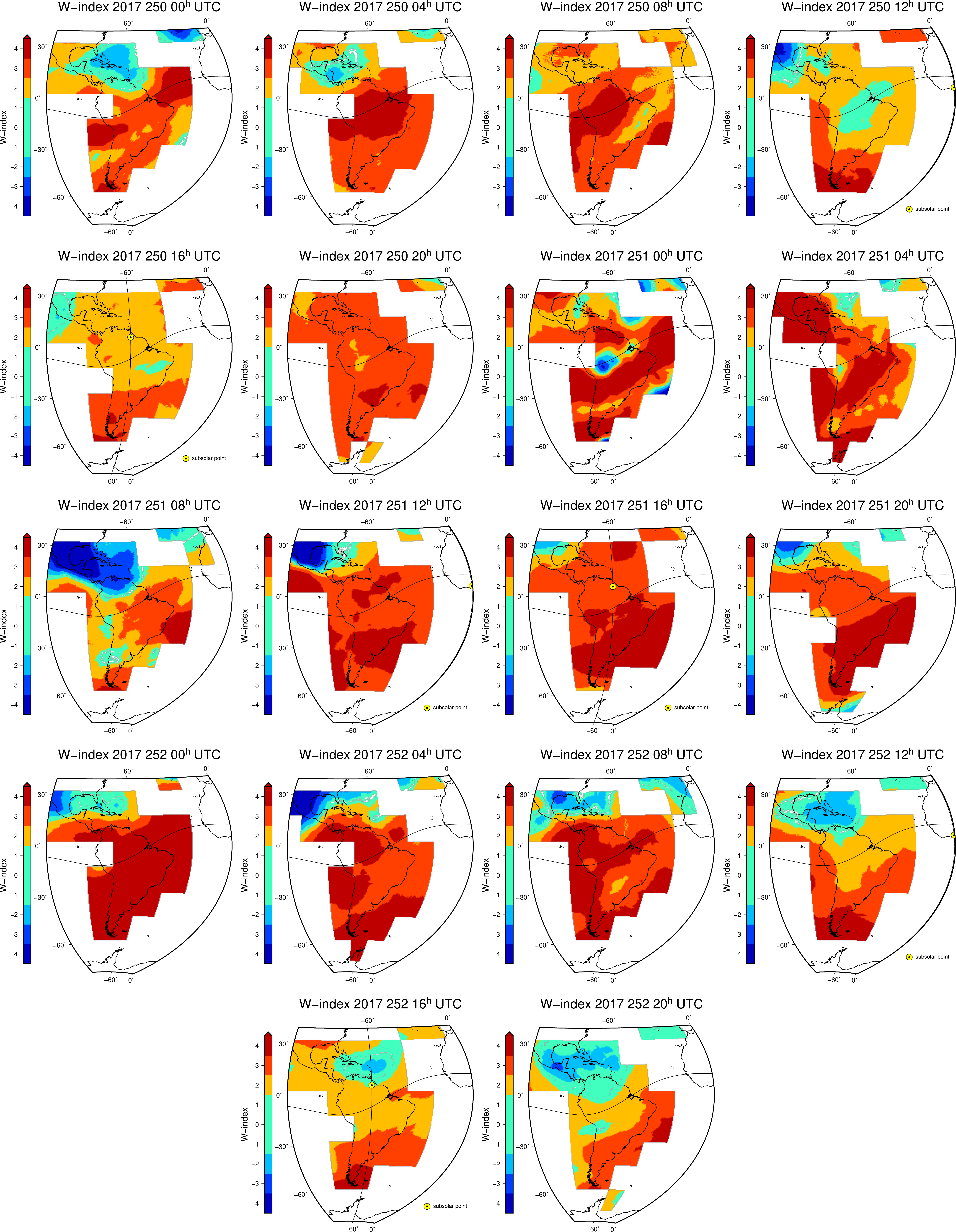}
\end{center}
\caption{Ionospheric W-index maps during the first CMEs associated with an X9.3-class solar flare on 6 September 2017, the largest solar X-ray flare seen in 12 years with multiple partial halo ejecta. They led to the sudden storm commencement SSC$=$23:44 UT of a severe geomagnetic storm.}
\label{MS_IS}
\end{figure*}

The performance of W-index for the quiet and moderate disturbed periods after the magnetic storm is presented in Fig.~\ref{QQ_QD}, whereas the geomagnetic indices are shown in Fig.~\ref{Dst_Kp}, where the selected interval is indicated between yellow arrows. The values of the W-index are mainly between -2 and 2, the positive perturbation prevailing towards negative geomagnetic latitude and equator and the negative disturbance becomes more evident towards positive latitudes.
Figure \ref{MS_IS} shows the evolution of two intense geomagnetic storms (marked between red arrows in Fig.~\ref{Dst_Kp}). The Dst value, which varies from $-15$ to $-20$ nT after the initial phase, reaches $-84$ nT at 23:00 UT (September 7th), $-125$ nT at 00:00 UT (same day) and $-142$ nT at 1:00 UT (September 8th). During the main phase of this event, again a severe storm has appeared with a minimum value of Dst $=-142$ nT at 15:00 UT on 8 September 2017. The W-index spatial distribution is very complex, as the responses of the ionosphere  to each consecutive storms is very different \citep{BLAGOVESHCHENSKY2019}. The first one is a ``classical'' storm, while the second is totally distinct. The mechanisms responsible for the two minimums in Dst, and the background of the spatial climate before each  Dst drop, are very different. The first storm was caused by solar wind perturbed by two consecutive shock waves of the CME, which is associated with an X9.3-class solar flare on 6 September 2017. The second storm was caused by the arrival of the second CME on September 8th.  The ionosphere before the first Dst minimum was affected by the intense solar flares, but still no geomagnetic storm occurred. The ionosphere before the second storm was already perturbed by the previous Dst minimum. Figure \ref{MS_IS} shows W-index values between +4 and $-3$ or $-4$ on September 8th at 00:00 UT (near the first minimum  Dst value), which are produced by the large increase of VTEC at low geomagnetic latitudes, and a remarkable decrease of VTEC at the Equator. The work by \citet{Paula2019} evidence the same results studying the ionospheric irregularity signatures on the SWARM-A electron density \citep[see Fig.~10 in][]{Paula2019}. Then, the Northern and Southern Hemisphere have different response to the recovery phase of the first storm, and the main phase of the second storm, highlighting large negative values of W-index at northern mid latitudes and large positive values at low and at southern mid latitudes. These characteristics almost disappear on September 9th at 12:00 UT. \citet{Gonzalez2018} and \citet{Imtiaz2019} show the same results using VTEC values from GNSS stations located at different latitudes in Central and South America. Gonzalez-Esparza et al. analyzed the VTEC values in Mexico. In particular, the GNSS station located at mid latitude showed a positive VTEC disturbance during the main phase of the first storm, whereas a negative disturbance was observed, at the same station, during the recuperation phase and also during the second geomagnetic storm.

%\textbf{\hspace{5mm}Quiet state and moderate disturbance}\par\medskip

% MAPAS TRANQUILOS

%% MAPAS DE TORMENTA

\section{Conclusions and Outlook}
A multi-GNSS, operational, high-rate and openly accessible ionospheric TEC monitoring system for South America has been successfully developed, tested and implemented. Both the comparison of the produced maps against global products, including several final IGS GIMs, and also against independent and highly accurate dSTEC observations, resulted in mean biases and RMS lower than 1 TECU, respectively.
Offline W-index maps generated from our VTEC operational products inherit their high temporal and spatial resolutions (15 min and $0.5 \times 0.5$ degrees, respectively), and they were capable to describes fairly well the variability of the ionosphere. These maps proved useful to analyze the main characteristics of a complex perturbation, both spatially and  temporally, and were in agreement with independent analyses. Accordingly, we are highly motivated to develop operational W-index maps, in parallel with our VTEC products, in order to improve the capabilities of the ionosphere monitoring system. We hope to achieve this goal in the near future.

\hspace{-0.7cm}\textbf{Data Availability Statement}

A plot of the most recent TEC map can be accessed anonymously from wilkilen
.fcaglp.unlp.edu.ar/ion/latest.png (or alternatively in Spanish from wilkilen
.fcaglp.unlp.edu.ar/ion/ultimo.png), whereas registered users can retrieve the
TEC maps produced by the system, in IONEX and NetCDF formats, from wilkilen
.fcaglp.unlp.edu.ar/ion/magn/.

\section*{Acknowledgements}
This research was supported by  the Agencia Nacional de Promoción 
Científica y Tecnológica (ANPCyT, Argentina), trough the program Fondo para la Investigación Científica y Tecnológica (FONCyT), with grants PICT 2015-3710, PICT-2015-1776 and by Universidad Nacional de La Plata with grant 11/G142. The authors thank the International GNSS Service (IGS) for providing the IONEX maps, and also the NASA/GSFC's Space Physics Data Facility's OMNIWeb Plus Service. We would like to thank the people, organizations and agencies responsible to collect,compute, maintain and openly provide the observations, products and databases that made this work possible. In particular, real-time GNSS orbit and clock messages,along with GNSS observations, are provided by the Bundesamt für Kartographie und Geodäsie (BKG, Germany), in support to the IGS. Real-time GNSS observations are also provided by: the Instituto Brasileiro de Geografia e Estatıstica (IBGE, Brazil), the Instituto Geográfico Nacional (IGN, Argentina), the National Aeronautics and Space Administration (NASA, USA), in support to the IGS, and the Instituto Geográfico Militar (IGM, Uruguay); also, this material is (partially) based on data services provided by the UNAVCO Facility with support from the National Science Foundation (NSF, USA) and NASA under NSF Cooperative Agreement No. EAR-0735156 (UNAVCO, USA).
%% Harvard
%\bibliographystyle{model2-names.bst}\biboptions{authoryear}
\bibliographystyle{elsarticle-harv}

\bibliography{manuscript}

\end{document}